\documentclass{aastex631}

\usepackage{graphicx}

\usepackage{multirow}
\usepackage{ mathrsfs }

\begin{document}

\title{Search for correlations of high-energy neutrinos detected in IceCube with radio-bright AGN and gamma-ray emission from blazars\footnote{Released on xx, 2022}}

\affiliation{III. Physikalisches Institut, RWTH Aachen University, D-52056 Aachen, Germany}
\affiliation{Department of Physics, University of Adelaide, Adelaide, 5005, Australia}
\affiliation{Dept. of Physics and Astronomy, University of Alaska Anchorage, 3211 Providence Dr., Anchorage, AK 99508, USA}
\affiliation{Dept. of Physics, University of Texas at Arlington, 502 Yates St., Science Hall Rm 108, Box 19059, Arlington, TX 76019, USA}
\affiliation{CTSPS, Clark-Atlanta University, Atlanta, GA 30314, USA}
\affiliation{School of Physics and Center for Relativistic Astrophysics, Georgia Institute of Technology, Atlanta, GA 30332, USA}
\affiliation{Dept. of Physics, Southern University, Baton Rouge, LA 70813, USA}
\affiliation{Dept. of Physics, University of California, Berkeley, CA 94720, USA}
\affiliation{Lawrence Berkeley National Laboratory, Berkeley, CA 94720, USA}
\affiliation{Institut f{\"u}r Physik, Humboldt-Universit{\"a}t zu Berlin, D-12489 Berlin, Germany}
\affiliation{Fakult{\"a}t f{\"u}r Physik {\&} Astronomie, Ruhr-Universit{\"a}t Bochum, D-44780 Bochum, Germany}
\affiliation{Universit{\'e} Libre de Bruxelles, Science Faculty CP230, B-1050 Brussels, Belgium}
\affiliation{Vrije Universiteit Brussel (VUB), Dienst ELEM, B-1050 Brussels, Belgium}
\affiliation{Department of Physics and Laboratory for Particle Physics and Cosmology, Harvard University, Cambridge, MA 02138, USA}
\affiliation{Dept. of Physics, Massachusetts Institute of Technology, Cambridge, MA 02139, USA}
\affiliation{Dept. of Physics and The International Center for Hadron Astrophysics, Chiba University, Chiba 263-8522, Japan}
\affiliation{Department of Physics, Loyola University Chicago, Chicago, IL 60660, USA}
\affiliation{Dept. of Physics and Astronomy, University of Canterbury, Private Bag 4800, Christchurch, New Zealand}
\affiliation{Dept. of Physics, University of Maryland, College Park, MD 20742, USA}
\affiliation{Dept. of Astronomy, Ohio State University, Columbus, OH 43210, USA}
\affiliation{Dept. of Physics and Center for Cosmology and Astro-Particle Physics, Ohio State University, Columbus, OH 43210, USA}
\affiliation{Niels Bohr Institute, University of Copenhagen, DK-2100 Copenhagen, Denmark}
\affiliation{Dept. of Physics, TU Dortmund University, D-44221 Dortmund, Germany}
\affiliation{Dept. of Physics and Astronomy, Michigan State University, East Lansing, MI 48824, USA}
\affiliation{Dept. of Physics, University of Alberta, Edmonton, Alberta, Canada T6G 2E1}
\affiliation{Erlangen Centre for Astroparticle Physics, Friedrich-Alexander-Universit{\"a}t Erlangen-N{\"u}rnberg, D-91058 Erlangen, Germany}
\affiliation{Physik-department, Technische Universit{\"a}t M{\"u}nchen, D-85748 Garching, Germany}
\affiliation{D{\'e}partement de physique nucl{\'e}aire et corpusculaire, Universit{\'e} de Gen{\`e}ve, CH-1211 Gen{\`e}ve, Switzerland}
\affiliation{Dept. of Physics and Astronomy, University of Gent, B-9000 Gent, Belgium}
\affiliation{Dept. of Physics and Astronomy, University of California, Irvine, CA 92697, USA}
\affiliation{Karlsruhe Institute of Technology, Institute for Astroparticle Physics, D-76021 Karlsruhe, Germany }
\affiliation{Karlsruhe Institute of Technology, Institute of Experimental Particle Physics, D-76021 Karlsruhe, Germany }
\affiliation{Dept. of Physics, Engineering Physics, and Astronomy, Queen's University, Kingston, ON K7L 3N6, Canada}
\affiliation{Department of Physics {\&} Astronomy, University of Nevada, Las Vegas, NV, 89154, USA}
\affiliation{Nevada Center for Astrophysics, University of Nevada, Las Vegas, NV 89154, USA}
\affiliation{Dept. of Physics and Astronomy, University of Kansas, Lawrence, KS 66045, USA}
\affiliation{Department of Physics and Astronomy, UCLA, Los Angeles, CA 90095, USA}
\affiliation{Centre for Cosmology, Particle Physics and Phenomenology - CP3, Universit{\'e} catholique de Louvain, Louvain-la-Neuve, Belgium}
\affiliation{Department of Physics, Mercer University, Macon, GA 31207-0001, USA}
\affiliation{Dept. of Astronomy, University of Wisconsin{\textendash}Madison, Madison, WI 53706, USA}
\affiliation{Dept. of Physics and Wisconsin IceCube Particle Astrophysics Center, University of Wisconsin{\textendash}Madison, Madison, WI 53706, USA}
\affiliation{Institute of Physics, University of Mainz, Staudinger Weg 7, D-55099 Mainz, Germany}
\affiliation{Department of Physics, Marquette University, Milwaukee, WI, 53201, USA}
\affiliation{Institut f{\"u}r Kernphysik, Westf{\"a}lische Wilhelms-Universit{\"a}t M{\"u}nster, D-48149 M{\"u}nster, Germany}
\affiliation{Bartol Research Institute and Dept. of Physics and Astronomy, University of Delaware, Newark, DE 19716, USA}
\affiliation{Dept. of Physics, Yale University, New Haven, CT 06520, USA}
\affiliation{Columbia Astrophysics and Nevis Laboratories, Columbia University, New York, NY 10027, USA}
\affiliation{Dept. of Physics, University of Oxford, Parks Road, Oxford OX1 3PU, UK}
\affiliation{Dipartimento di Fisica e Astronomia Galileo Galilei, Universit{\`a} Degli Studi di Padova, 35122 Padova PD, Italy}
\affiliation{Dept. of Physics, Drexel University, 3141 Chestnut Street, Philadelphia, PA 19104, USA}
\affiliation{Physics Department, South Dakota School of Mines and Technology, Rapid City, SD 57701, USA}
\affiliation{Dept. of Physics, University of Wisconsin, River Falls, WI 54022, USA}
\affiliation{Dept. of Physics and Astronomy, University of Rochester, Rochester, NY 14627, USA}
\affiliation{Department of Physics and Astronomy, University of Utah, Salt Lake City, UT 84112, USA}
\affiliation{Oskar Klein Centre and Dept. of Physics, Stockholm University, SE-10691 Stockholm, Sweden}
\affiliation{Dept. of Physics and Astronomy, Stony Brook University, Stony Brook, NY 11794-3800, USA}
\affiliation{Dept. of Physics, Sungkyunkwan University, Suwon 16419, Korea}
\affiliation{Institute of Physics, Academia Sinica, Taipei, 11529, Taiwan}
\affiliation{Dept. of Physics and Astronomy, University of Alabama, Tuscaloosa, AL 35487, USA}
\affiliation{Dept. of Astronomy and Astrophysics, Pennsylvania State University, University Park, PA 16802, USA}
\affiliation{Dept. of Physics, Pennsylvania State University, University Park, PA 16802, USA}
\affiliation{Dept. of Physics and Astronomy, Uppsala University, Box 516, S-75120 Uppsala, Sweden}
\affiliation{Dept. of Physics, University of Wuppertal, D-42119 Wuppertal, Germany}
\affiliation{Deutsches Elektronen-Synchrotron DESY, Platanenallee 6, 15738 Zeuthen, Germany }

\author[0000-0001-6141-4205]{R. Abbasi}
\affiliation{Department of Physics, Loyola University Chicago, Chicago, IL 60660, USA}

\author[0000-0001-8952-588X]{M. Ackermann}
\affiliation{Deutsches Elektronen-Synchrotron DESY, Platanenallee 6, 15738 Zeuthen, Germany }

\author{J. Adams}
\affiliation{Dept. of Physics and Astronomy, University of Canterbury, Private Bag 4800, Christchurch, New Zealand}

\author[0000-0002-9714-8866]{S. K. Agarwalla}
\altaffiliation{also at Institute of Physics, Sachivalaya Marg, Sainik School Post, Bhubaneswar 751005, India}
\affiliation{Dept. of Physics and Wisconsin IceCube Particle Astrophysics Center, University of Wisconsin{\textendash}Madison, Madison, WI 53706, USA}

\author[0000-0003-2252-9514]{J. A. Aguilar}
\affiliation{Universit{\'e} Libre de Bruxelles, Science Faculty CP230, B-1050 Brussels, Belgium}

\author[0000-0003-0709-5631]{M. Ahlers}
\affiliation{Niels Bohr Institute, University of Copenhagen, DK-2100 Copenhagen, Denmark}

\author[0000-0002-9534-9189]{J.M. Alameddine}
\affiliation{Dept. of Physics, TU Dortmund University, D-44221 Dortmund, Germany}

\author{N. M. Amin}
\affiliation{Bartol Research Institute and Dept. of Physics and Astronomy, University of Delaware, Newark, DE 19716, USA}

\author{K. Andeen}
\affiliation{Department of Physics, Marquette University, Milwaukee, WI, 53201, USA}

\author[0000-0003-2039-4724]{G. Anton}
\affiliation{Erlangen Centre for Astroparticle Physics, Friedrich-Alexander-Universit{\"a}t Erlangen-N{\"u}rnberg, D-91058 Erlangen, Germany}

% ...

\date{\today}

\collaboration{394}{IceCube Collaboration}

\begin{abstract}

The IceCube Neutrino Observatory sends realtime neutrino alerts with high probability of being astrophysical in origin. We present a new method to correlate these events and possible candidate sources using $2,089$ blazars from the \textit{Fermi}-LAT 4LAC-DR2 catalog and with $3,413$ AGNs from the Radio Fundamental Catalog. No statistically significant neutrino emission was found in any of the catalog searches. The result is compatible with a small fraction, $<1$\%, of AGNs being neutrino emitters and prior evidence for neutrino emission presented by IceCube and other authors from sources such as TXS~0506+056 and PKS~1502+06. We also present cross-checks to other analyses that claim a significant correlation using similar data samples, and we find that adding more information on the neutrino events and more data overall makes the result compatible with background. 

\end{abstract}


\section{Introduction} \label{sec:intro}

Astrophysical neutrinos are weakly-interacting, electrically neutral particles capable of revealing the origin of cosmic rays. Thought to be produced in regions near some of the most high-energy events in the Universe, astrophysical neutrinos travel undeflected by magnetic fields delivering information relating to their original production mechanism. One example of these regions is NGC~1068, an Active Galactic Nucleus (AGN), from which evidence of high-energy neutrino emission was recently found with the IceCube Neutrino Observatory~\citep{ngc1068}.

Despite ongoing research, the origins of astrophysical neutrinos remain largely unknown. However, AGNs are among the most promising candidate sources~\citep{ngc1068, fedeAGN, MuraseAGNS}. In particular, blazars -AGNs that have jets pointing towards Earth- are thought to be likely sources. Recently, evidence has emerged of a spatial and temporal correlation with the emission from blazar TXS~0506+056 and the high-energy neutrino event IC170922A, detected by IceCube~\citep{txspaper}. An alert was sent to other observatories and a follow-up campaign in multiple wavelengths was started. At the moment of neutrino detection, the blazar was flaring in gamma rays, and it was determined that the chance of accidental coincidence was at the 3$\sigma$ level. Subsequent retrospective analyses from the IceCube collaboration found an excess in neutrino events with respect to the expected atmospheric background coincident with the position of TXS~0506+056, with a significance at the 3.5$\sigma$ level~\citep{txsnuflare}. Later studies have also indicated a possible correlation between the blazar PKS~1502+106 and the IceCube neutrino event IC190730A~\citep{Rodrigues_2021}.

Both IC170922A and IC190730A were high-energy neutrino events with a high probability of being astrophysical in origin. For each of these events, IceCube released public realtime ``neutrino alerts''. For an alert to be issued, the event must have passed at least one of several filtering streams based on its reconstructed energy and direction~\citep{Aartsen:2016lmt, Blaufuss:2019fgv}. These alerts can be used to search for correlations between different candidate sources, as other studies have shown, e.g. a search for correlations with radio-selected AGNs in~\cite{Plavin_2020} and with blazars in~\cite{Giommi_2020}. 

In this paper we develop a method to perform correlation analyses using the latest catalog of alerts that IceCube has published, IceCat-1~\citep{icecat1}\footnote{Data release in \url{https://dataverse.harvard.edu/dataset.xhtml?persistentId=doi:10.7910/DVN/SCRUCD}}, and various catalogs of candidate sources. These analyses also use different weighting schemes for the sources to account for different neutrino production mechanisms. As a use case, we present three analyses: two using blazars from \textit{Fermi}~4LAC-DR2 catalog and one with AGNs from the RFC 2022a catalog. The paper is organized as follows: in Section~\ref{sec:icecube} the IceCube Neutrino Observatory is introduced, in Section~\ref{sec:samples} the description of the neutrino catalog and the two source catalogs considered for this paper are discussed and in Section~\ref{sec:method} we introduce the analysis technique. In Section~\ref{sec:results} the results are shown. Section~\ref{sec:discussion} contains the conclusion and a comparison to previous analyses that used similar samples.

\section{The IceCube Neutrino Observatory}\label{sec:icecube}

The IceCube Neutrino Observatory is a cubic-kilometer detector located at the South Pole~\citep{Aartsen_2017}. It is a three-dimensional array of $5,160$~Digital Optical Modules (DOMs) deployed along strings between depths of $1,450\,$m and $2,450\,$m below the ice. Each DOM contains a photomultiplier tube (PMT)~\citep{ABBASI2010139} and readout electronics~\citep{Abbasi_2009}. The array of DOMs detects neutrinos indirectly via the Cherenkov radiation produced by the relativistic charged particles that result from the interactions of neutrinos in the ice or bedrock below IceCube. The analyses presented in this paper are focused on track-like events, produced by charged-current muon-neutrino interactions.

\section{Data samples}\label{sec:samples}

\subsection{IceCube's neutrino alert catalog}

The IceCube Realtime System was established in 2016~\citep{Aartsen:2016lmt}. Recently, the selection criteria were updated~\citep{Blaufuss:2019fgv} and now consists of two alert streams, ``Gold'' and ``Bronze'' neutrino alerts, which are defined by an average astrophysical purity (\textit{signalness}) of 50\% and 30\% respectively. 
The signalness is calculated as the ratio between the expected number of astrophysical signal events and the expected total number of signal and background events with energy larger than the energy of the event and at the given declination. The expected number of signal events is calculated assuming a spectral index of $2.19\pm0.10$~\citep{Haack:2017E1} for the signal neutrino flux. 
The latest diffuse astrophysical muon neutrino spectrum measured by IceCube has an index of $2.37\pm0.09$~\citep{diffuse9.5}, but the signalness definition is kept unchanged for consistency. Typically, alert events have reconstructed neutrino energies above $100\,$TeV. 

When IceCube detects a high-energy neutrino that passes the alert criteria, an automatic GCN\footnote{General Coordinates Network, \url{https://gcn.nasa.gov}} Notice is sent out to other observatories to allow for follow-up observations in multiple wavelength bands. This first notice contains the reconstructed direction with statistical uncertainties calculated with a fast reconstruction algorithm~\citep{abbasi2021muontrack}. Another, more sophisticated reconstruction algorithm is then run and a second, updated GCN Notice and a GCN Circular are sent out a few hours later, including the refined reconstructed direction and uncertainties that also include systematic uncertainties.

\begin{figure}[!ht]
    \centering
    \includegraphics[trim={0 0 0 1.8cm},clip,width=0.6\textwidth]{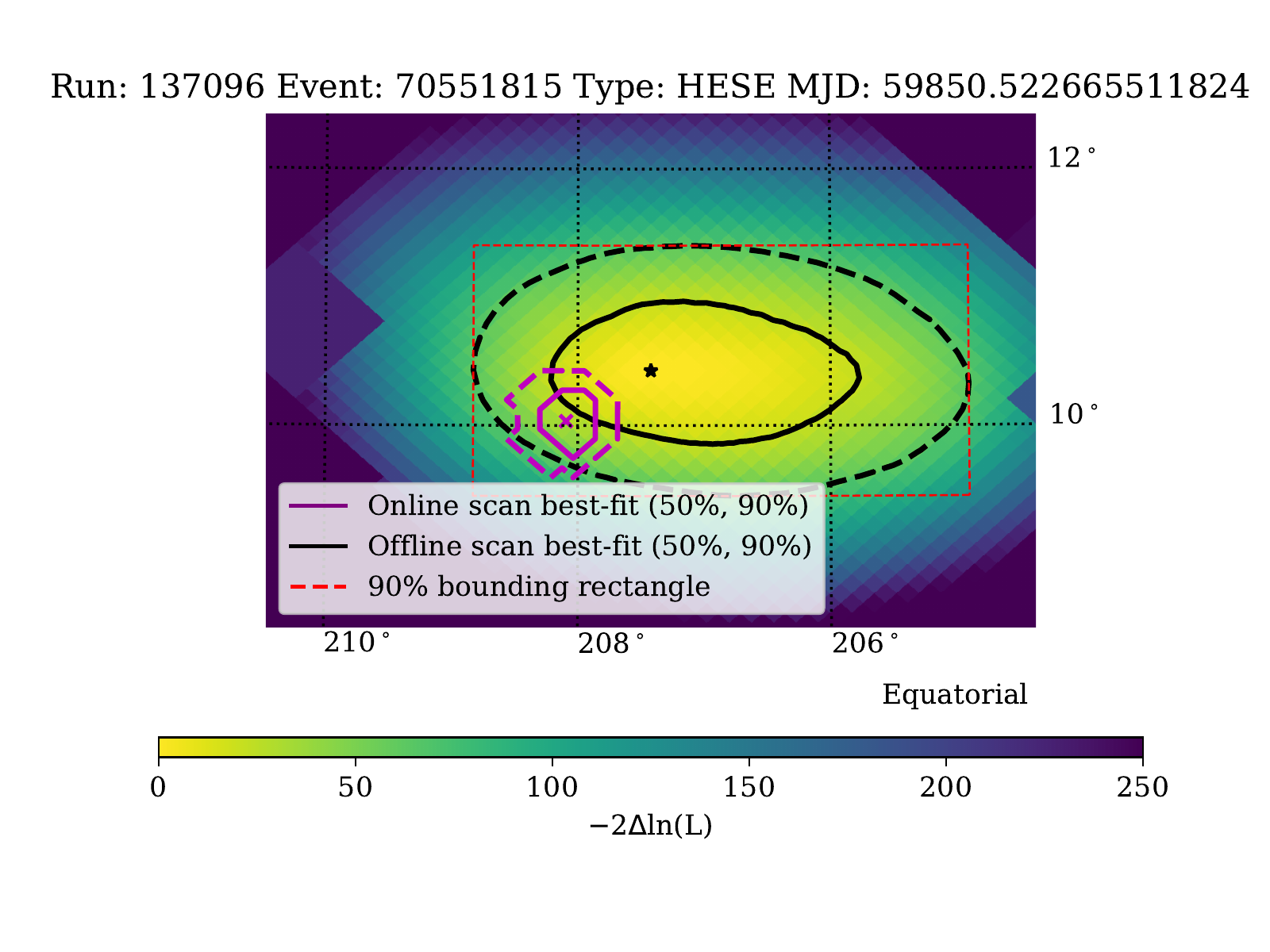} 
    \caption{Example of a log-likelihood map of a neutrino alert reconstructed with the full-sky maximum likelihood reconstruction algorithm in right ascension and declination. The black star is the best-fit position from the likelihood scan. The solid and dashed black curves represent the 50\% and 90\% confidence level respectively, derived from the likelihood scan. The red dashed rectangle determines the errors quoted in the GCN Circulars and is defined as the minimum rectangle that would encapsulate the uncertainty contours. Lastly, the purple solid and dashed curves are the 50\% and 90\% confidence level contours calculated with the initial, less refined reconstruction method and sent out in the first GCN Notice.}
    \label{fig:exampleLLH}
\end{figure}

This second algorithm is a maximum likelihood reconstruction method~\citep{energyreco} that determines the neutrino direction by running the track reconstruction for each direction $(\theta, \phi)$ on a HEALPix~\citep{Gorski_2005} grid covering the sky, with iteratively finer binning. The likelihood is calculated at every point of the grid, and the direction maximizing the likelihood represents the track whose light yield matches the observed data and therefore is used as the reconstructed direction of the neutrino. This likelihood map can be normalized to act as a spatial probability density function for the neutrino (Figure~\ref{fig:exampleLLH}). From the map one can also derive uncertainty contours following the procedure explained in~\cite{myicrc}, in which an event is simulated and reconstructed multiple times to determine the appropriate confidence level and to account for systematic uncertainties. In practice, all the events published in the realtime alert system in IceCube are calibrated using the simulations of one archival event, IC160427A. The work presented in~\cite{myicrc} indicates that this method has difficulties in correctly accounting for systematic uncertainties and leads to a possible overcoverage of the angular resolution. 

\begin{figure}[!ht]
    \centering
    \includegraphics[trim={0 0 0 0.6cm},clip, width=0.6\textwidth]{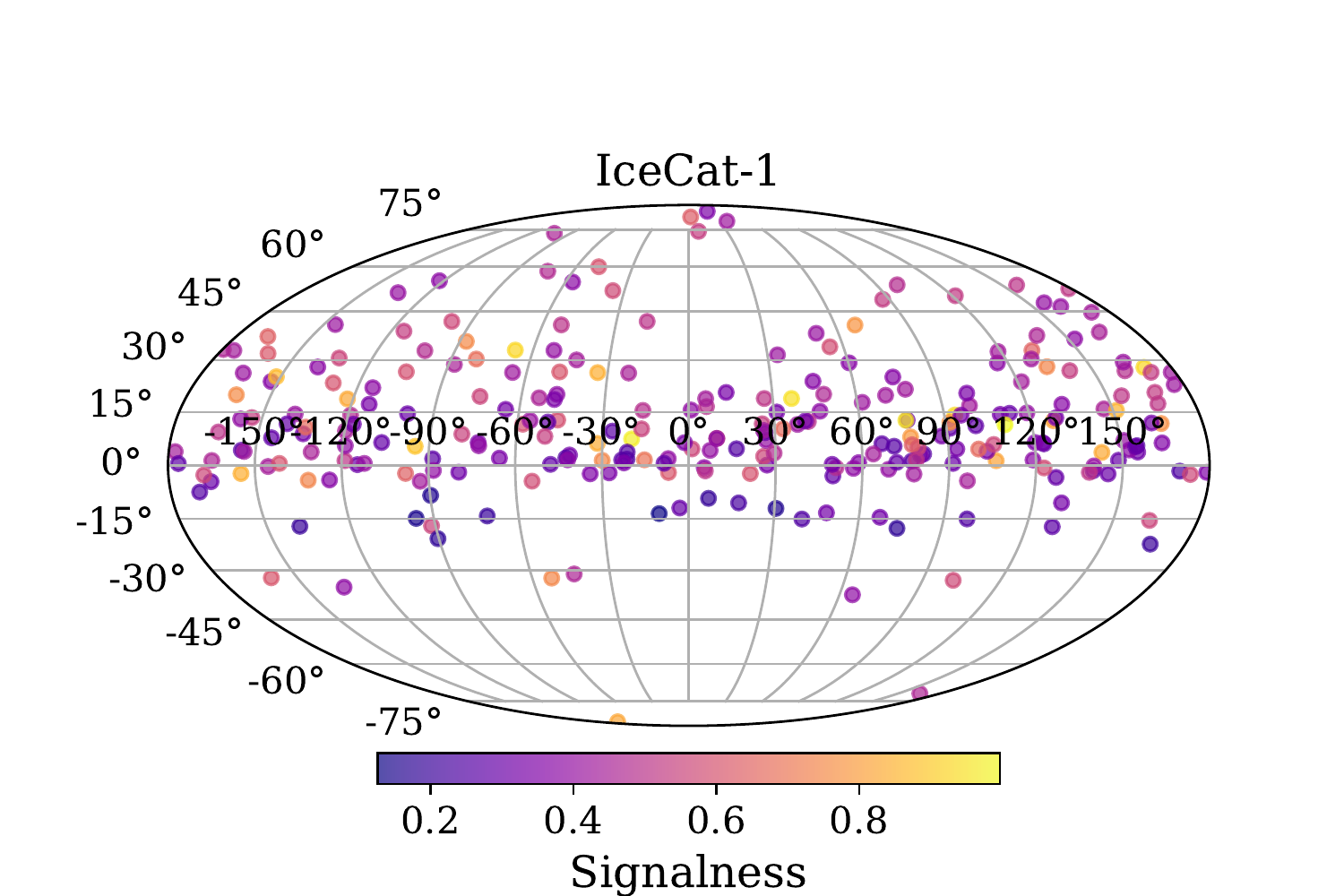} 
    \caption{Neutrino events which have passed the alert criteria in the period between 2011 and 2020, in equatorial coordinates. The colours represent the probability that the neutrino is of astrophysical origin (signalness). The varying density in the number of events reflects the dependence of IceCube's effective area on the declination of the neutrino direction.}
    \label{fig:v2cat}
\end{figure}

The IceCube Event Catalog of Alert Tracks-1 (IceCat-1)~\citep{icecat1} includes all the alerts that were issued with the revised criteria as well as archival data starting from 2011, prior to the start of the realtime stream, which would have been selected as realtime alerts and were processed in the same way. In total, it contains 275 track-like events. \cite{icecat1} also shows a simple correlation analysis searching for spatial coincidences between neutrino alerts and different candidate sources.  The correlation analysis presented here represents an updated method with increased sensitivity to potential signal, since it includes additional information of the alert event and the candidate counterpart. In Figure~\ref{fig:v2cat}, the spatial distribution of the alerts is shown, with the colours representing the signalness of each event.

\subsection{Fermi-LAT sources}

For the first two analyses, blazars were selected from the incremental version of the Fourth Catalog of Active Galactic Nuclei detected by \textit{Fermi}-LAT (4LAC-DR2, for Data Release 2)~\citep{4lac, dr2}. This catalog contains all AGNs detected by \textit{Fermi}-LAT between 2008 and 2018 at high galactic latitudes $b$ ($|b| > 10\,$deg). We select blazars (BL-Lacs, Flat Spectrum Radio Quasars and blazar candidates of unknown types), which represent 98\% of the catalog. 

The catalog was further refined by studying the distribution of sources in the sky to assure that it is isotropic, which fulfills the condition of isotropy that goes into the definition of the Test Statistic, explained in Section~\ref{sec:method}. Various selection criteria with increasing limits on the brightness were applied to the gamma-ray energy flux $F$, which is defined as the energy flux of photons of a certain energy per unit area, $F = (\mathrm{d}N/\mathrm{d}E\mathrm{d}A\mathrm{d}t) E^{2}$, and obtained by spectral fitting from $100\,$MeV to $100\,$GeV. The latitude and longitude distributions of the blazar data after the selection were compared to a uniform distribution, both weighted by the energy flux and unweighted. A cut of $ F \geq 10^{-11.6}\,$erg~cm$^{-2}$~s$^{-1}$ was chosen based on the results retaining 90\% of the cumulative flux from all objects in the catalog. The final number of blazars used in the analysis is $2,089$ (from the original $2,863$ objects in 4LAC-DR2).

Two weighting schemes are used in analyzing the 4LAC-DR2 catalog. For the first, the normalized 10-year average energy flux in gamma rays above $100\,$MeV of each blazar is used as the weight, since the blazars with higher flux are expected to emit more high-energy neutrinos~\citep{PhysRevD.90.023007, Petropoulou:2015upa} and, therefore, are more likely to be the source of an alert neutrino. A similar weighting scheme was followed in a previous IceCube analysis using a different neutrino catalog~\citep{Huber:20192z}. The second weighting scheme utilizes the light curves from the Light Curve Repository\footnote{\url{https://fermi.gsfc.nasa.gov/ssc/data/access/lat/LightCurveRepository/}} made available by the \textit{Fermi}-LAT collaboration~\citep{LCRATel}. Here we use the normalized flux in the 1-month time bin that coincides with the neutrino arrival time as the weight, and thereby use the blazar's temporal information. This procedure is similar to the one followed in \cite{txspaper}. From the $2,089$ blazars selected for this analysis, $1,161$ had flux variability and were released in the light curve repository. The weights are calculated as follows:

\begin{itemize}
    \item If the light curve of a blazar is not provided in the repository, it is assumed that it is not variable and the average energy flux is considered as the flux at the time of the neutrino's arrival.
    
    \item If there is a flux measurement at the neutrino arrival time, it is used as the weight.
    
    \item If there is no flux measurement in the time bin of the neutrino arrival or it is an upper limit, we check the monthly time bin before and after the time bin of the neutrino event (neighbouring bins). If both have a flux measurement, the weight is an average of both. If only one of them has a measurement, the weight is equal to that flux. This assumes that the monthly energy flux does not vary greatly from one month to another.
    
    \item If there is no flux measurement or only an upper limit is provided in the time bin of the neutrino event and its neighbouring time bins, then a weight of zero is assigned. The weight is set to zero also when there are no measurements available in the neighbouring bins. This is to avoid mixing the hypotheses for the two weighting schemes.
    
\end{itemize}

\subsection{Radio Fundamental Catalog sources}

The second catalog we consider for coincidences with IceCat-1 is the Radio Fundamental Catalog (RFC), in its 2022a version\footnote{\url{http://astrogeo.org/sol/rfc/rfc_2022a/}}. It contains $20,499$ objects detected between 1980-04-11 and 2022-03-24 at $22\,$GHz, $8.6\,$GHz, $5\,$GHz and $2.2\,$GHz observed with Very Long Baseline Interferometry (VLBI) observations~\citep{10.1093/mnras/118.3.276}. 

For the analysis presented here, we use the flux density integrated over VLBI images at $8.6\,$GHz as a weight and select AGNs with flux density $S_{8.6\,\mathrm{GHz}} \geq 150\,$mJy, which leaves a total of $3,413$ AGNs (828 in common with the subsample from 4LAC-DR2 described in the previous section). We follow the selection criteria applied in~\cite{Plavin_2020} with the purpose of cross-checking the results obtained in their analysis. The procedure followed to compare both methods is explained in Appendix~\ref{sec:appendix} and discussed in Section~\ref{sec:discussion}.

\section{Analysis technique} \label{sec:method}

The statistical significance of the overall correlation between the neutrino alerts and various candidate source catalogs is calculated using the stacking analysis method. In this method, the hypothesis that we test is that the neutrino is correlated with a coincident source from the candidate source class. ``Coincident'' is defined here as the source lying within the 90\% error contour of the reconstructed arrival direction of the neutrino. The null hypothesis is that the neutrino is not correlated with any sources, either because it is atmospheric background or because it comes from a source not included in the catalog being tested. The likelihood $\mathscr{L}$ is the weighted sum of a signal Probability Density Function (PDF), $S$, and a background PDF, $B$, 

\begin{equation}
    \mathscr{L} = \frac{n_S}{N}S + \left(1 -\frac{n_S}{N}\right) B 
\end{equation}
where $n_S$ is the number of signal events and $N$ is the total number of signal and background neutrino candidate events. The Test Statistic (TS) for a single neutrino $i$ ($N = 1$) is defined as

\begin{equation}\label{eq:ts}
    \mathrm{TS}_i = -2 \log \left( \frac{\mathscr{L}(n_S = 1)}{\mathscr{L}(n_S  = 0)}\right) = - 2 \log \frac{\hat{S}}{B}.
\end{equation}

Treating each neutrino independently, the hypothesis is tested for each coincident source $b$. The combination returning the highest TS is selected as the most likely counterpart. If TS~$<0$, the background hypothesis is chosen, i.e. $S/B = 1$. The signal PDF is defined as

\begin{equation}\label{eq:signal}
    S = S_{\mathrm{signalness},\nu}\cdot S_{\mathrm{spatial}, \nu}(\vec{x_b})\cdot w_b,
\end{equation}
where $S_{\mathrm{signalness},\nu}$ is the signalness of the neutrino alert; $S_{\mathrm{spatial}, \nu}(\vec{x_b})$ is the neutrino point-spread function, calculated using the re-scaled likelihood maps from the reconstruction of the alerts that depends on the position of the source $\vec{x_b}$ (see Section~\ref{sec:samples}); and $w_{b}$ is the source weight that depends on the hypothesis of neutrino production mechanism. The most likely counterpart is then the source for which $S_{\mathrm{spatial}, \nu}(\vec{x_b})\cdot w_b$ is maximum, and we denote that in Equation~\ref{eq:ts} as $\hat{S}$. The signalness is a fixed parameter for each individual neutrino, so it does not affect the selection of the counterpart.

For the background case, in which the neutrino is not associated with any source, the probability of finding a correlation by chance can be assumed to be isotropic, and therefore the background PDF is simply defined by $B = B_{\mathrm{spatial}} = \frac{1}{4\pi}$. Finally, the TS of the data is calculated by summing over all neutrino alerts, TS$_{\mathrm{data}} = \sum_{i} \mathrm{TS}_i$. To calculate the significance of the result, N = 5000 background pseudo-experiments are constructed by fixing the direction of the neutrino alerts and scrambling the candidate sources in right ascension and declination, while maintaining the high galactic latitude criterion from the \textit{Fermi} 4LAC-DR2 catalog of $|b|>10\,$deg. Then the TS is calculated for each of the background maps and its distribution is fitted to a gamma distribution for the $p$-value estimation. 

For the analysis using \textit{Fermi}-LAT 4LAC catalog with average flux weights, the sensitivity flux, which is defined as neutrino flux needed in order to get a TS larger than the median of the background distribution in 90\% of the cases, is at 5.1\% of the astrophysical flux, which corresponds to 6.3 signal correlations from the 275 neutrinos in the IceCat-1 catalog, i.e. 6.3 neutrinos correlated with their sources. The sensitivity for the same catalog with the monthly energy flux weighting scheme (second analysis) is at 4.4\% of the astrophysical flux, which is equivalent to 5.5 signal correlations. Using the RFC catalog and the average VLBI the sensitivity flux is 6\% of the astrophysical flux, or 7.4 signal correlations.

\section{Results}\label{sec:results}

\begin{figure}[!ht]

  \includegraphics[width=0.33\linewidth]{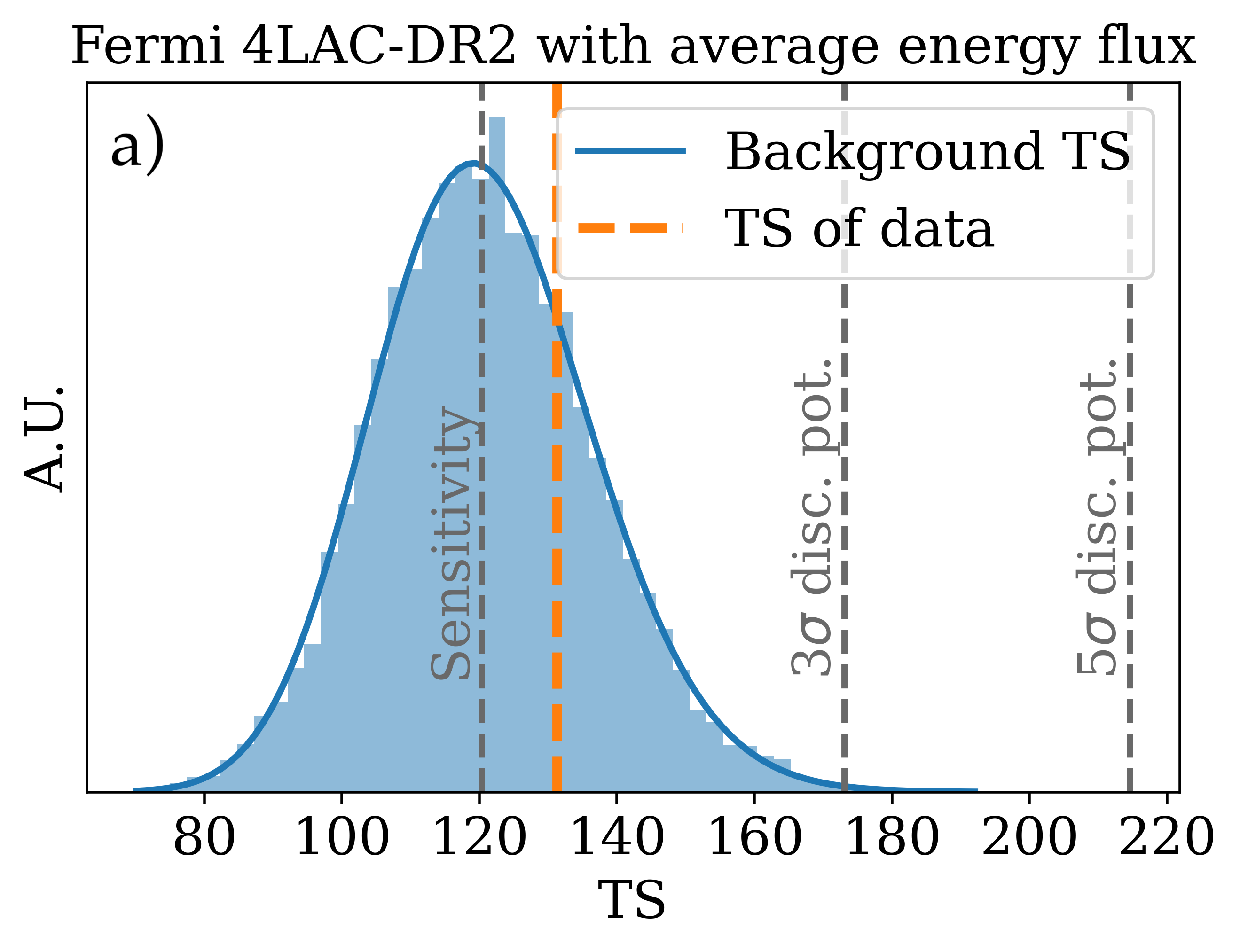}
  \includegraphics[width=0.33\linewidth]{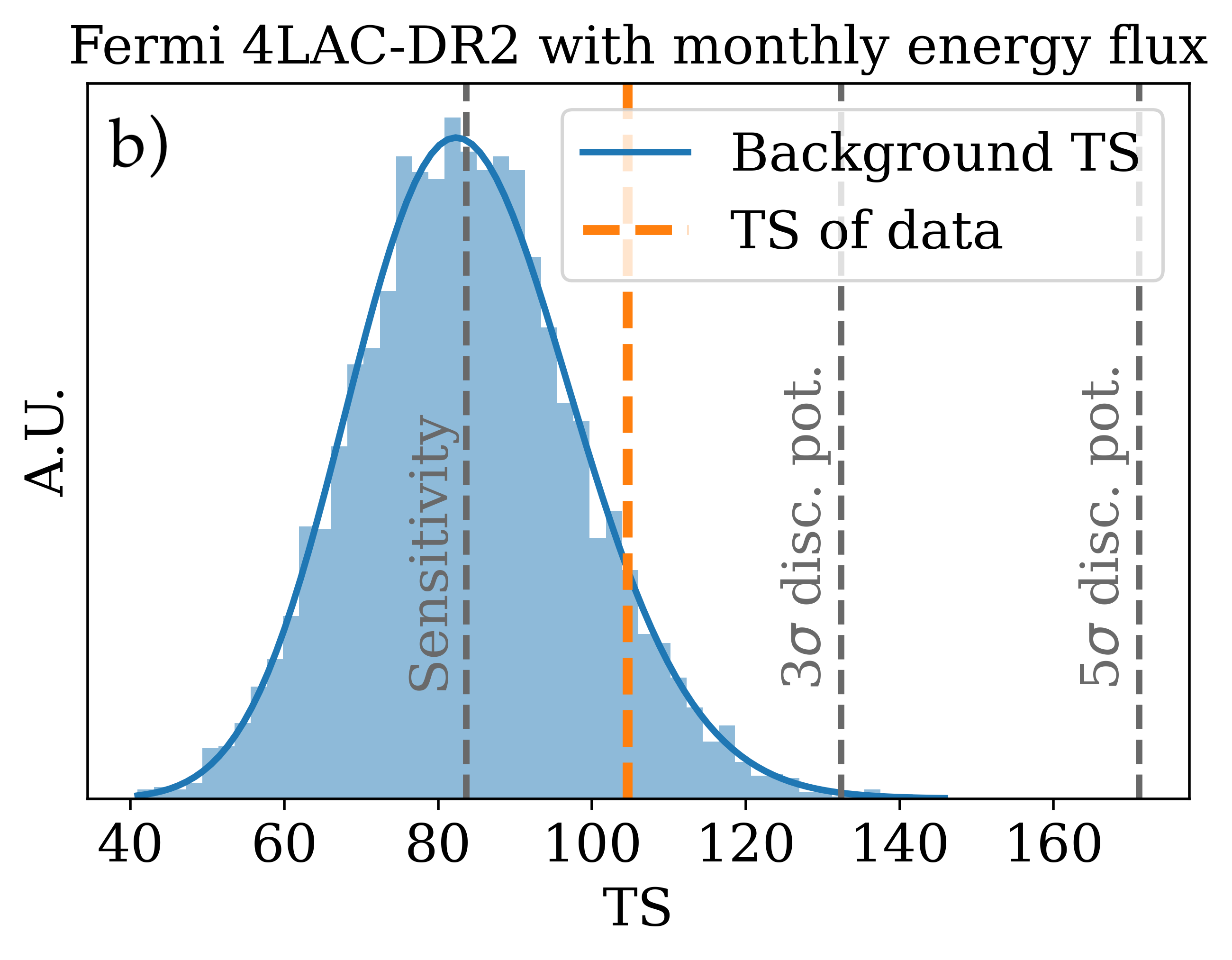}
  \includegraphics[width=0.33\linewidth]{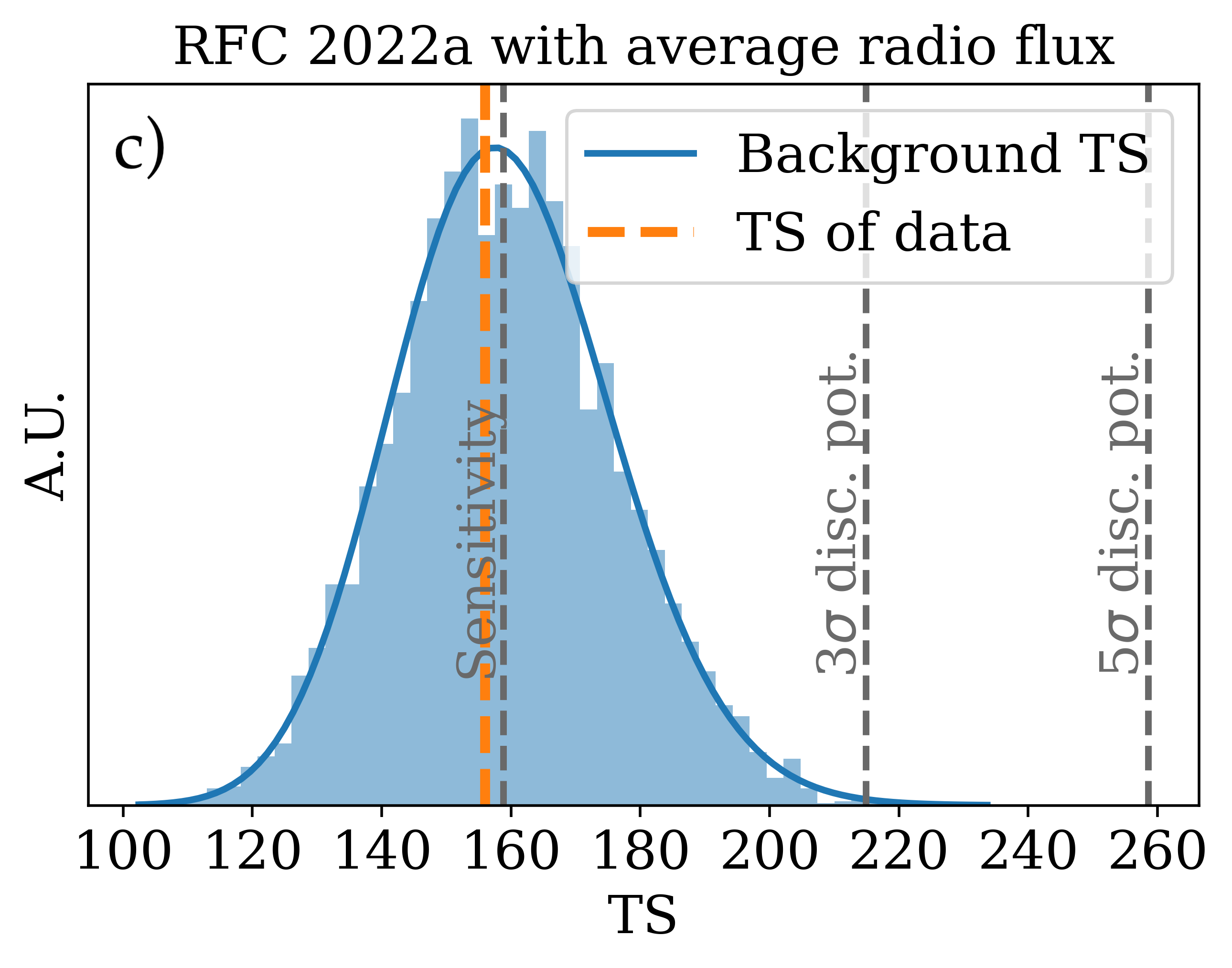}
  
\caption{The blue histogram shows the distribution of the TS values for $N = 5000$ background maps, fitted to a gamma distribution (blue curve). The orange dotted line represents the observed TS of the data, while the grey lines represent the TS$_{\text{data}}$ needed to reject the background hypothesis at the $3\sigma$ or $5\sigma$ level. Panel a) For the \textit{Fermi} 4LAC-DR2 catalog with the average energy flux weights,  TS$_{\text{data}}$~=~131.3, while TS~=~120.4 for the median background maps, leading to a $p$-value of 0.248. Panel b) For the \textit{Fermi} 4LAC-DR2 catalog with the monthly energy flux weights, TS$_{\text{data}}$~=~104.6, median background maps has TS~=~83.6 and the $p$-value is 0.08. Panel c) For the RFC 2022a catalog, TS$_{\text{data}}$~=~155.8, TS~=~158.9 for the median background maps and the $p$-value is 0.57.}
\label{fig:tsbkg}
\end{figure}

\begin{table}[!htb]
    \caption{The table shows the 10 correlations that contribute most to the TS of the data using the \textit{Fermi} 4LAC-DR2 catalog and average flux weights. The first column is the neutrino using the nomenclature of the realtime program, the second column is the signalness of each neutrino, the third column is the J2000 name of the most likely counterpart, the fourth column is the angular distance between the candidate source and the neutrino (calculated as the angle between the two sightlines) and the last column is the contribution of each correlation to the final TS.}
    \centering
        \begin{tabular}{c|c|c|c|c}
            Neutrino  & Signalness & Blazar              & $d$ [deg] & TS   \\ \hline
            IC170922A & 63 \%      & TXS 0506+056        & 0.11           & 7.43 \\
            IC190730A & 67 \%      & PKS 1502+106        & 0.27           & 7.13 \\
            IC120523A & 49 \%      & 3C 454.3            & 0.72           & 7.12 \\
            IC111208A & 45 \%      & Mkn 421             & 0.78           & 5.63 \\
            IC190704A & 49 \%      & NVSS J104516+275136 & 1.04           & 5.14 \\
            IC190504A & 39 \%      & NVSS J042025-374443 & 0.50           & 4.18 \\
            IC200109A & 77 \%      & TXS 1100+122        & 0.36           & 4.05 \\
            IC181014A & 39 \%      & PKS 1454-354        & 0.99           & 3.87 \\
            IC150104A & 45 \%      & MG2 J180948+2910    & 0.50           & 3.41 \\
            IC130127A & 61 \%      & PKS B2330-017       & 0.58           & 3.17
        \end{tabular}
        \label{tab:fermi-average}
\end{table}

\begin{table}[!htb]
    \caption{The table shows the 10 correlations that contribute most to the TS of the data using the \textit{Fermi} 4LAC-DR2 catalog and monthly flux weights. The first column is the neutrino using the nomenclature of the realtime program, the second column is the signalness of each neutrino, the third column is the J2000 name of the most likely counterpart, the fourth column is the angular distance between the candidate source and the neutrino (calculated as the angle between the two sightlines) and the last column is the contribution of each correlation to the final TS.}

    \centering

        \begin{tabular}{c|c|c|c|c}
            Neutrino  & Signalness & Blazar              & $d$ [deg] & TS   \\ \hline
            IC170922A & 63 \%      & TXS 0506+056        & 0.11           & 9.43 \\
            IC190730A & 67 \%      & PKS 1502+106        & 0.27           & 5.63 \\
            IC111208A & 45 \%      & Mkn 421             & 0.78           & 5.36 \\
            IC130125A & 53 \%      & S5 0016+73          & 1.03           & 4.87 \\
            IC190504A & 39 \%      & NVSS J042025-374443 & 0.50           & 4.41 \\
            IC150104A & 49 \%      & MG2 J180948+2910    & 0.50           & 4.28 \\
            IC111216A & 95 \%      & TXS 0222+185        & 0.51           & 4.13 \\
            IC200109A & 78 \%      & TXS 1100+122        & 0.36           & 3.96 \\
            IC140101A & 56 \%      & PMN J1243-0218      & 1.33           & 3.81 \\
            IC181014A & 39 \%      & PKS 1454-354        & 0.99           & 3.54
        \end{tabular}
        \label{tab:fermi-monthly}
\end{table}

\subsection{Fermi 4LAC-DR2 with average energy flux}

No significant correlation was found between the $2,089$ selected blazars from \textit{Fermi} 4LAC-DR2 weighted by the average 10-year energy flux and the neutrino alerts, with a $p$-value of 0.248, see Figure~\ref{fig:tsbkg} Panel a. The 10 most significant correlations (the 10 blazar-neutrino pairs with the highest TS$_i$ in the data) are listed in Table~\ref{tab:fermi-average}. The two strongest correlations have been extensively studied, (see Section~\ref{sec:intro}, TXS~0506+056 - IC170922A and PKS~1502+106 - IC190730A). The third and fourth correlation are also interesting since 3C~454.3 and Mkn~421 are two of the brightest sources in the sky, which explains the high TS, but in both cases the area of the error region of the associated neutrino is very large. The areas of the 90\% uncertainty contours are $26.80\,$deg$^2$ and $56.07\,$deg$^2$ respectively, compared to the median 90\% contour area of all alerts which is $5.96\,$deg$^2$, or the 90\% contour area of IC170922A and IC190730A, of $1.54\,$deg$^2$ and $5.61\,$deg$^2$, respectively.

\subsection{Fermi 4LAC-DR2 with monthly energy flux} 

The TS is again consistent with the null hypothesis when the selected blazars are weighted by the energy flux at the neutrino arrival time (see Figure~\ref{fig:tsbkg} Panel b). The $p$-value of this analysis is 0.08. Table~\ref{tab:fermi-monthly} shows the 10 most contributing correlations to the TS. The main contribution to TS$_{\mathrm{data}}$ is given by TXS~0506+056 and IC170922A, since the blazar was flaring in gamma rays at the neutrino arrival time as mention in Section~\ref{sec:intro}, leading to a higher TS value than observed in the analysis with average energy flux weights. On the other hand, one can see that the blazar 3C 454.3 was in a quiescent state at the neutrino arrival time, and therefore, the TS value of that correlation is low and is not within the 10 most contributing ones.

\subsection{RFC 2022 with average VLBI flux} 

The result from this third analysis, shown in Figure~\ref{fig:tsbkg} Panel c, using the RFC catalog and the average VLBI flux at $8.6\,$GHz as weight for the contribution of the AGNs is that the data is compatible with the null hypothesis, with a $p$-value of 0.57. In Table~\ref{tab:rfc} the 10 correlations with the highest TS are shown. It is worth noting that the correlations of TXS~0506+056 with IC170922A and PKS~1502+106 with IC190730A appear on the list (with the nomenclature used in the RFC catalog for the AGNs), since they are both very bright in radio as well.

\begin{table}[]
\centering
\caption{The table shows the 10 correlations that contribute most to the TS of the data using the RFC 2022a catalog. The first column is the neutrino using the nomenclature of the realtime program, the second column is the signalness of each neutrino, the third column is the J2000 name of the most likely AGN counterpart, the fourth column is the angular distance between the candidate source and the neutrino and the last column is the contribution of each correlation to the final TS.}
\begin{tabular}{c|c|c|c|c}
Neutrino  & Signalness & AGN & $d$ [deg] & TS   \\ \hline
IC120523A & 49 \%      & J2253+1608 & 0.72 & 5.62 \\
IC170922A & 63 \%      & J0509+0541 & 0.12 & 5.41 \\
IC190730A & 67 \%      & J1504+1029 & 0.28 & 5.21 \\
IC150812A & 83 \%      & J2151+0552 & 0.47 & 4.12 \\
IC141110A & 29 \%      & J1652+0618 & 0.26 & 3.51 \\
IC150428A & 32 \%      & J0204+1514 & 0.44 & 3.48 \\
IC130125A & 53 \%      & J0019+7327 & 1.02 & 3.19 \\
IC170422A & 39 \%      & J1550+0527 & 3.29 & 3.05 \\
IC140307A & 40 \%      & J2025+3343 & 1.67 & 3.02 \\
IC200530A & 59 \%      & J1659+2629 & 0.48 & 2.99
\end{tabular}
\label{tab:rfc}
\end{table}

\section{Discussion and conclusions}\label{sec:discussion}

In this paper we have described a new method developed within IceCube that allows us to calculate the overall correlation between neutrino alerts and a catalog of candidate sources. We conducted three analyses: two searching for correlations with blazars from \textit{Fermi} 4LAC-DR2 catalog and one with AGNs from the RFC 2022a catalog. A significant correlation was not found in any of the three cases, meaning that the majority of the neutrino alerts are neither produced in bright or flaring gamma-ray blazars nor in radio-bright AGNs. 

The analyses presented here, and the method itself, rely heavily on the robustness of the reconstruction. Due to the way the contours are calculated, which determines how broad the neutrino spatial PDF is, a better handle on the systematic uncertainties is crucial. The reconstruction method used in IceCube's realtime system is being improved and further modifications will be implemented in the future~\citep{myicrc}, which might give an opportunity to revisit this analysis with improved sensitivity. 

The results obtained in this paper are, however, compatible with previous hints of a signal, e.g.~\cite{txspaper, Rodrigues_2021}. The expected number of chance correlations that have $TS_{i}> 7$ is 0.077 for the 4LAC-DR2 catalog weighted by the average energy flux. In data, there are three such correlations, which point to potential signal being present but non discernible from chance coincidences. The observed $TS_{\mathrm{data}}$ is compatible with the TS distribution that one would obtain if only 1\% of the blazars in the selected \textit{Fermi} 4LAC-DR2 catalog emits neutrinos. This is calculated by randomly choosing 1\% of sources in multiple pseudo-experiments and injecting them as signal, i.e., the location of the blazar is moved into the error region of a neutrino so that a correlation is forced. The same study was performed with RFC 2022a, with similar results. This indicates the possibility that only a small fraction of the gamma-ray bright and radio-bright AGN population are sources of high-energy neutrinos.

The third correlation analysis was designed to be similar to~\cite{Plavin_2020} and allows for comparison. Adding more neutrino events and additional information, such as the signalness or the spatial PDF from the reconstruction, leads to a TS$_{data}$ compatible with background. Even when the definition of ``coincidence'' resembles the one in their paper (i.e., when the spatial term of the signal PDF is simplified to a uniform distribution inside of the error contour), we do not observe the same significance. The details of these cross-checks can be found in Appendix~\ref{sec:appendix}. 

The possibility of testing several hypotheses for the search of correlations with neutrino realtime alerts is the main strength of the method that was developed and explained in this paper. The information from the realtime reconstruction algorithm and the signalness of the alert allow us to be more sensitive to potential signal, and they were used for the first time in an analysis for the results presented here.

\section*{Acknowledgements}
The IceCube collaboration acknowledges the significant contribution to this manuscript from Cristina Lagunas Gualda. The authors gratefully acknowledge the support from the following agencies and institutions: 
USA {\textendash} U.S. National Science Foundation-Office of Polar Programs,
U.S. National Science Foundation-Physics Division,
U.S. National Science Foundation-EPSCoR,
Wisconsin Alumni Research Foundation,
Center for High Throughput Computing (CHTC) at the University of Wisconsin{\textendash}Madison,
Open Science Grid (OSG),
Advanced Cyberinfrastructure Coordination Ecosystem: Services {\&} Support (ACCESS),
Frontera computing project at the Texas Advanced Computing Center,
U.S. Department of Energy-National Energy Research Scientific Computing Center,
Particle astrophysics research computing center at the University of Maryland,
Institute for Cyber-Enabled Research at Michigan State University,
and Astroparticle physics computational facility at Marquette University;
Belgium {\textendash} Funds for Scientific Research (FRS-FNRS and FWO),
FWO Odysseus and Big Science programmes,
and Belgian Federal Science Policy Office (Belspo);
Germany {\textendash} Bundesministerium f{\"u}r Bildung und Forschung (BMBF),
Deutsche Forschungsgemeinschaft (DFG),
Helmholtz Alliance for Astroparticle Physics (HAP),
Initiative and Networking Fund of the Helmholtz Association,
Deutsches Elektronen Synchrotron (DESY),
and High Performance Computing cluster of the RWTH Aachen;
Sweden {\textendash} Swedish Research Council,
Swedish Polar Research Secretariat,
Swedish National Infrastructure for Computing (SNIC),
and Knut and Alice Wallenberg Foundation;
European Union {\textendash} EGI Advanced Computing for research;
Australia {\textendash} Australian Research Council;
Canada {\textendash} Natural Sciences and Engineering Research Council of Canada,
Calcul Qu{\'e}bec, Compute Ontario, Canada Foundation for Innovation, WestGrid, and Compute Canada;
Denmark {\textendash} Villum Fonden, Carlsberg Foundation, and European Commission;
New Zealand {\textendash} Marsden Fund;
Japan {\textendash} Japan Society for Promotion of Science (JSPS)
and Institute for Global Prominent Research (IGPR) of Chiba University;
Korea {\textendash} National Research Foundation of Korea (NRF);
Switzerland {\textendash} Swiss National Science Foundation (SNSF);
United Kingdom {\textendash} Department of Physics, University of Oxford.

\appendix

\section{Comparisons to published analyses} \label{sec:appendix}

\cite{Plavin_2020} find a $3\sigma$ correlation between 56 IceCube high-energy neutrinos and a previous version of the Astrogeo catalog\footnote{\url{http://astrogeo.org/vlbi/solutions/rfc_2019c/}}. Using the same selection criteria for the candidate sources, we repeat this analysis with a few changes to the method, including more of the available information. 

One of the key differences is the definition of the TS, which in~\cite{Plavin_2020} analysis is simply the geometric average of the VLBI flux of the sources coincident with the neutrinos. Their neutrino selection consists of alert and alert-like events from different catalogs published by IceCube, and a table with the full catalog is included along with the paper. They are chosen so that only high-energy ($E>200\,$TeV) tracks with good angular resolution ($\Omega_{90} < 10\,$deg$^2$) are included in the catalog. 56 events from 2009 to 2019 pass their criteria, 44 of which are also contained in the IceCat-1 catalog. The remaining 12 events are from the period from 2009 to 2011 that was not included in the IceCat-1 catalog (10) or didn't pass our current alert criteria (2). 

\begin{figure}[!ht]
    \centering
    \includegraphics[trim={0 4.5cm 0 4.5cm},clip,width=0.9\textwidth]{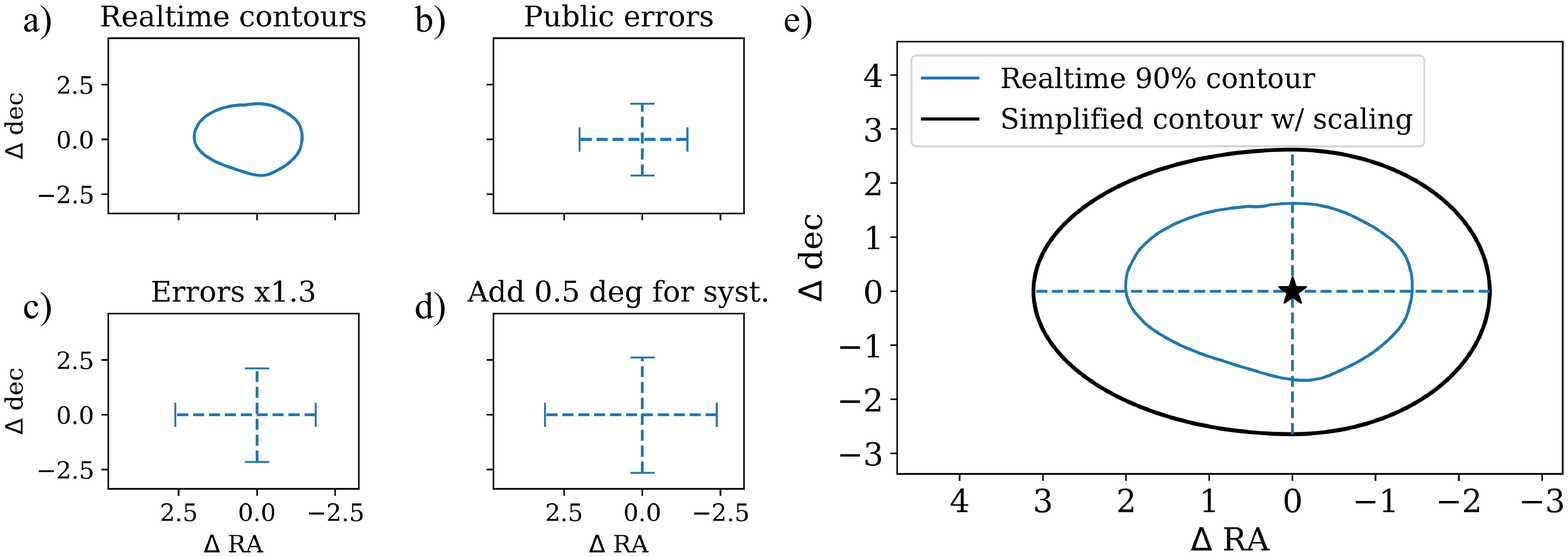} 
    \caption{Example of the simplification and scaling that is applied to the realtime contours in~\cite{Plavin_2020}, which is also done in the cross check presented here. From the original contours in Panel a), only the minimum rectangle is published, which is translated to uncertainty in the coordinates (Panel b). Those errors are multiplied by 1.3 (Panel c) and an extra parameter is added to account for systematic uncertainties, leading to a minimum $p$-value at an added 0.5 degrees (Panel d). The error bars are joined by elliptical segments to obtain the simplified contours corresponding to the best-fit position from the reconstruction, represented by the black star (Panel e).}
    \label{fig:plavin}
\end{figure}

For comparison, our definition of the TS is kept, but the spatial term in the signal PDF is calculated as in \cite{Plavin_2020}. The information that IceCube publishes in the GCN Circular whenever an alert is detected is the reconstructed direction along with the minimum rectangle that encapsulates the 90\% error contour in right ascension and declination (see Figure~\ref{fig:exampleLLH}). In~\cite{Plavin_2020} the coordinate-wise errors are multiplied by the ratio of the 90\% quantiles of 2D and 1D Gaussian distributions ($\approx$1.3). An extra parameter $x$ is added linearly in all directions to account for possible unknown systematic uncertainties. It varies from 0 to 1~deg in increments of 0.1~degrees and fitted to find the minimum pre-trial $p$-value, which is obtained at $x = 0.5\,$deg in~\cite{Plavin_2020}. Finally, the error region is assumed to be composed of 4 points joint by elliptical segments. This is shown in Figure~\ref{fig:plavin}. 

IceCube already calculates the errors by assuming a 2D distribution, but the extra 1.3 factor is also applied in the comparison for consistency with~\cite{Plavin_2020}. Systematic uncertainties are added to the contours before publishing them, although one could argue that there are unknown systematics. This is currently being studied, see e.g.~\cite{myicrc}, but the method to calibrate the errors is conservative and the real uncertainties could be smaller. However, it is worth investigating this scaling up of the contours since other analyses have also found interesting hints of signal following a similar procedure (although in some cases with a very different neutrino data sample focused on lower energies), e.g.,~\cite{Giommi_2020, Buson_2022}.

For the comparison analysis we switch the spatial term of the signal PDF from the neutrino point-spread function (based on the reconstruction's likelihood landscape) to the simpler coincidence-based function described above, multiplying the angular errors by 1.3 and considering the added 0.5~deg as a fixed value. Then the $S_{\mathrm{spatial}}$ is 1 for all AGNs inside the area of the calculated contour and 0 outside. This is performed with the datasets described in Section~\ref{sec:samples} (IceCat-1 catalog + RFC 2022a) and the original dataset from~\cite{Plavin_2020} (56 alert and alert-like events + RFC 2019c) with and without the scaling factors, 1.3 from the quantiles + 0.5 degrees for systematic uncertainties. Without scaling, the error bars would be similar to the ones in Figure~\ref{fig:plavin} Panel b), while for the scaled-up contours the error bars would be as in Figure~\ref{fig:plavin} Panel d). For the neutrinos from the 56 alert and alert-like event catalog that do not have a signalness, a signalness of 45.1\% is assumed, which is the average of the IceCat-1 catalog.

The results can be seen in Table~\ref{tab:comparison-plavin}. The lowest $p$-value is achieved with the test that is the most similar to~\cite{Plavin_2020} (same AGN and neutrino catalogs and scaling of the angular uncertainty). However, when more events are added, the significance drops. The same happens when the angular errors are not scaled up, which indicates that the correlations which are contributing to the TS are from AGNs that are from outside the original contours, and therefore far away from the best-fit position.

\begin{table}[]
\centering
\caption{Results from the cross-checks to \cite{Plavin_2020}. }
\begin{tabular}{|c|c|c|c|}
\hline
\multicolumn{2}{|c|}{Check}                                         & $p$-value & Significance \\ \hline
\multirow{2}{*}{RFC 2022a + IceCat-1}         & With scaling    & 0.048   & 1.67         \\ 
                                                & Without scaling & 0.32    & 0.47         \\ \hline
\multirow{2}{*}{RFC 2019c + Alert-like events} & With scaling    & 0.0043  & 2.63         \\
                                                & Without scaling & 0.2     & 0.84        \\\hline
\end{tabular}
\label{tab:comparison-plavin}
\end{table}

Only one of the four correlations that drive the result in~\cite{Plavin_2020} appears in Table~\ref{tab:rfc}, and in the two tests with the RFC 2019c + alert-like events selection, due to its high signalness and the brightness of the AGN. The four most significant correlations from~\cite{Plavin_2020} are found in the test of RFC 2019c + alert-like events including scaling as expected, since this is the test that is closest to their analysis. With the information available, we cannot confirm the findings of~\cite{Plavin_2020}, and the only hint of excess disappears when we use a more sophisticated description of the spatial probability density function for the neutrino events.

\bibliography{main}{}
\bibliographystyle{aasjournal}

\end{document}